\documentclass[aps,pre,english,twocolumn,showpacs,preprintnumbers,amsmath,amssymb,floatfix,superscriptaddress,longbibliography]{revtex4-1}

\usepackage[T1]{fontenc}
\usepackage[latin9]{inputenc}
\usepackage{graphicx}
\usepackage{amssymb}
\usepackage{babel}
\makeatletter

\usepackage[version=3]{mhchem}

\usepackage{color}

\global\arraycolsep=2pt
\usepackage{stmaryrd}
\usepackage{amsmath}
\usepackage{amssymb}
\usepackage{graphicx}
\usepackage{textcomp}
\usepackage{calrsfs}
\usepackage{yfonts}
\usepackage{bm}
\usepackage{color}
\newcommand{\ben}{\begin{equation*}}
\newcommand{\een}{\end{equation*}}
\newcommand{\bean}{\begin{eqnarray*}}
\newcommand{\eean}{\end{eqnarray*}}

\newcommand{\be}{\begin{equation}}
\newcommand{\ee}{\end{equation}}
\newcommand{\bea}{\begin{eqnarray}}
\newcommand{\eea}{\end{eqnarray}}

\newcommand{\psumbar}{\sum\nolimits^\prime}
\DeclareMathOperator*{\psum}{\psumbar}

\makeatother
\usepackage{babel}
\makeatother

\usepackage{color}

\begin{document}

%\title{Partial melting induced growth of atmospheric ice particles}

\title{Secondary ice growth mechanism for ice nuclei in the atmosphere}

\author{M.  Bostr{\"o}m}
\email{mathias.bostrom@ensemble3.eu}
\affiliation{Centre of Excellence ENSEMBLE3 Sp. z o. o., Wolczynska Str. 133, 01-919, Warsaw, Poland}

\author{Y. Li}
 \email{leon@ncu.edu.cn}
  \affiliation{Department of Physics, Nanchang University, Nanchang 330031, China}
  \affiliation{Institute of Space Science and Technology, Nanchang University, Nanchang 330031, China}

\author{I. Brevik}
%\email{iver.h.brevik@gmail.com}
\affiliation{Department of Energy and Process Engineering, Norwegian University of Science and Technology, NO-7491 Trondheim, Norway}

\author{C. Persson}
%\email{clas.persson@fys.uio.no}
\affiliation{Department of Materials Science and Engineering, KTH Royal Institute of Technology, SE-100 44 Stockholm, Sweden}
\affiliation{Centre for Materials Science and Nanotechnology, Department of Physics, University of Oslo, P. O. Box 1048 Blindern, NO-0316 Oslo, Norway}

\author{S. Carretero-Palacios}
%  \email{sol.carretero@uam.es}
\affiliation{Departamento de F\'isica de Materiales and Instituto de Ciencia de Materiales Nicol\'as Cabrera, Universidad Aut\'onoma de Madrid, 28049 Madrid, Spain}

\author{O. I. Malyi}
%\email{oleksandr.malyi@ensemble3.eu}
\affiliation{Centre of Excellence ENSEMBLE3 Sp. z o. o., Wolczynska Str. 133, 01-919, Warsaw, Poland}

\begin{abstract}
%It may seem counterintuitive that partial internal melting of an atmospheric ice nucleus can actually lead to further ice growth. However, quantum vacuum fluctuation-induced Casimir-Lifshitz interaction is seen to promote ice growth on the outer surface of partially melted ice particles (treated as a mixture of solid ice and liquid water). %
The study of atmospheric ice nuclei is vital for understanding the formation of precipitation and the development of cloud systems as it reveals how these tiny particles grow. A mechanism of such growth when the nuclei are in a mixed ice/water phase and quantum vacuum fluctuation-induced Casimir-Lifshitz interaction highlights the complexity and interconnection of the atmospheric processes with quantum theory. Initially of the order of $0.1\sim10\rm\mu m$ in size, atmospheric ice nuclei can expand by the accumulation of water molecules from the surrounding water vapor.
%The study of atmospheric ice nuclei is vital for understanding the formation of precipitation and the development of cloud systems as it reveals how these tiny particles grow and evolve. A mechanism of such growth through partial internal melting (i.e. in a mixed phase of ice and water) and quantum
%\textcolor{blue}{and thermal? in the main text it is also written thermal fluctuations}  vacuum fluctuation-induced Casimir-Lifshitz interaction highlights the complexity and interconnectedness of the atmospheric processes with quantum theory.  Initially of the order $0.1\sim10\rm\mu m$ in size, atmospheric ice nuclei can increase its size by the accumulation of water molecules from the surrounding water vapor. }
\end{abstract}
\date{\today}

\maketitle

\subsection*{Introduction}

\par Small ice particles form in clouds when the water vapor freezes onto an ice nucleation particle (INP) that can be volcanic ash, soot, desert dust, organic matter, bacteria, pollen, and fungal spores, amongst others. The condensation on such nucleus promotes the growth of ice crystals. The exact mechanisms for deposition are not fully understood yet~\cite{ResurgenceinIceNucleiMeasurementResearch,Korolev_acp-22-13103-2022}, despite several proposed theories, e.g., Wegener-Bergeron-Findeisen~\cite{findeisen2015,StorelvmoTan2015} process. This process, due to the thermodynamic instability of water-ice mixtures, turns liquid clouds into ice clouds. The water vapor at lower moisture concentrations accumulates more readily on ice than on water~\cite{StorelvmoTan2015}. This influences, via changes in optical properties,  not only weather but also climate~\cite{StorelvmoTan2015}. The importance of very small ice particles on ice-phase processes for the global hydrological cycle and for precipitation, is known~\cite{MulenstadtSourdevalDalanoeQuaasGRL2015}. As discussed by Schrod et al.~\cite{Schrod_etal_acp-20-12459-2020}, there are a number of unresolved questions regarding the nature of ice nuclei: ``their geographical and vertical distribution, seasonal variation, and the types of aerosol particles that contribute to their population even in today's atmosphere''. The initial ice nuclei, known to be of the order $0.1\sim1\rm\mu m$~\cite{ValiIceNucleai_Nature1966}, seem to require specific surface sites to induce heterogeneous ice growth. For example, certain feldspars have highly specific surface sites that seem to play vital roles~\cite{atkinson2013importance,Murraydoi:10.1126/science.aam5320,kiselev2017active}. This kind of heterogeneous ice formation, occurring for instance in tropical cumulus clouds over sea, can appear at relatively high temperatures ranging between 0 to $-5\rm\ ^\circ\rm C$~\cite{Lloydetalacp-20-3895-2020}. Observations of much higher concentrations of ice particles than expected were offered with various potential explanations by Lloyd et al.~\cite{Lloydetalacp-20-3895-2020}: (i) presence of more efficient ice nucleating particles, (ii) recycling of ice within the downwelling mantle of the convective cloud,  and finally (iii) through a secondary ice production process. Different secondary ice production mechanisms have been discussed in the past\,\cite{Korolev_acp-22-13103-2022}, including (1) fragmentation in connection with droplet freezing, (2) the Hallett-Mossop process (rime-splintering mechanism), (3) fragmentation via ice-ice collision, (4)  thermal shock induced ice particle fragmentation, (5) fragmentation of sublimating ice, and (6) the activation of ice-nucleating particles in transient supersaturation around freezing drops. We will here propose one additional, previously missing, secondary ice production process.

%\begin{figure}
%  \centering
% \includegraphics[width=1.0\columnwidth]{Figures/Cartoon.jpg}
%  \caption{\label{fig:Cartoon}  \textcolor{blue}{} }
%\end{figure}

\par Homogeneous ice nucleation occurs at very low temperatures. In this letter, we propose a secondary ice production process that can be either heterogeneous or homogeneous, depending on initial conditions. This ice crystal growth starts from a nucleus consisting of a micron-sized partially melted ice particle. It should be made clear that we do not address the nature of the initial seeding, which brings about this ice particle. As demonstrated, at temperatures as high as the freezing point of water, the Casimir-Lifshitz free energy, and thus stress~\cite{Dzya}, owing to the quantum and thermal fluctuations, cause the accumulation of ice at the interface towards water vapor, which can lead to thicknesses of the order of nanometers to micrometers, depending on both the initial size and the degree of partial melting of the ice nucleus. This corresponds to a substantial volume expansion of the ice particles, easily exceeding $40\%$ for nucleus of diameters around $1-10\rm\ \mu m$. Our model is initially inspired by the pioneering work by Elbaum and Schick~\cite{Elbaum}, which explored the impacts of Casimir-Lifshitz interaction in ice/water systems. This milestone paper~\cite{Elbaum} stimulated a lasting interest in the scientific community on the premelting of ice and its effects and consequences~\cite{Dash1995,Wettlaufer,benet16,benet19,li2019interfacial,Prachi2019role}. Nevertheless, several groups~\cite{JohannesWater2019,LUENGOMARQUEZMacDowell2021,LuengoMarquez_IzquierdoRuiz_MacDowell2022,LiMiltonBrevikMalyiThiyamPerssonParsonsBostrom_PRB2022} have shown that improved models for the dielectric functions for ice and water give very different results compared to the original work. These new results have been applied to the silver iodide as ice nucleation particles in atmospheric sciences~\cite{LUENGOMARQUEZMacDowell2021}, the astrogeophysics~\cite{BostromEstesoFiedlerBrevikBuhmannPerssonCarreteroParsonsCorkery2021}, and the anomalous stabilization of methane gas hydrates inside pores~\cite{LiCorkeryCarreteroBerlandEstesoFiedlerMiltonBrevikBostrom2023}. The growth of thin ice layers, instead of liquid water films, on a partially melted ice-vapor interface is easily seen from textbook arguments for Lifshitz-free energy, where the intermediate layer grows when it has dielectric properties in between the surrounding media~\cite{Dzya}.  While no water films are induced, we mention that recent works~\cite{LuengoMarquez_IzquierdoRuiz_MacDowell2022} indicate how an atomically thin water layer can form at the ice-vapor interface.
% (check, 2023-02-01)

\subsection*{Theory considerations at the triple point of water}

\begin{figure}
  \centering
  \includegraphics[width=1.0\columnwidth]{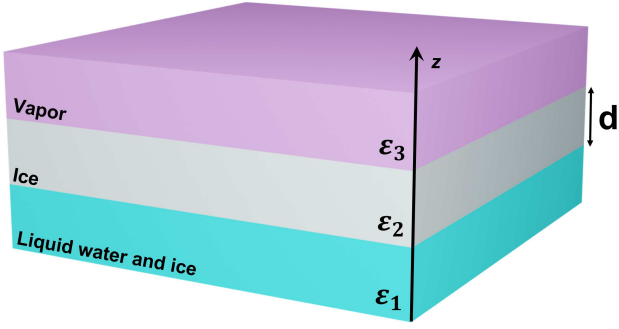}
  \caption{\label{Fig.SchematicFig1} (Color online) Schematic diagram of the three-layer planar model system with a mixture of ice and water, $\varepsilon_1$; pure \ce{H2O} ice, $\varepsilon_2$; and vapor, $\varepsilon_3$. }
\end{figure}

\par Parashar et al.~\cite{Prachi2019role} demonstrated that the variation of the Casimir-Lifshitz energy, excluding the self-energy part, can be computed accurately for layered concentric spherical ice particles. According to that work, to a good approximation, micron-sized particles could be treated as quasi-planar from a theoretical point of view. At the very least, it gives the right order of magnitude for the grown ice layer thickness. The three-layer system, namely the partially melted ice-ice-vapor, gives a reasonable estimate of the contributions of the quantum and thermal fluctuations that drive the secondary growth mechanism here proposed. Some additional important effects, not included in this letter, are Laplace pressures~\cite{Marcolliacp-20-3209-2020}, the role of pressures on ice nucleation rates~\cite{EspinosaPhysRevLett.117.135702}, etc. Although we mainly discuss the growth of ice layers on partially melted ice cores in terms of planar layered systems, the limitations for micron-sized ice cores are briefly outlined.

\par We model the dielectric function of the ice particles with a volume averaged theory~\cite{NAVID20084159}, employing the imaginary parts of complex dielectric functions of water ($\varepsilon_w''$) and ice ($\varepsilon_i''$) as input for the Kramers-Kronig relationship\,\cite{landau2013statistical}
\begin{eqnarray}
\varepsilon_1(i \zeta_m)
&=&
1+\frac{2}{\pi}\int_0^\infty d\omega \frac{\Phi \omega  \varepsilon_w''(\omega)+(1-\Phi) \omega  \varepsilon_i''(\omega)}{\omega^2+\zeta_m^2}
\nonumber\\
&=&
\Phi \varepsilon_w(i \zeta_m)+(1-\Phi) \varepsilon_i(i \zeta_m),
\end{eqnarray}
%\begin{eqnarray}
%\varepsilon_1(i \zeta_m)
%&=&
%\frac{2}{\pi}\int_0^\infty d\omega \frac{\Phi \zeta_m \varepsilon_w'(\omega)+(1-\Phi) \zeta_m \varepsilon_i'(\omega)}{\omega^2+\zeta_m^2}
%\nonumber\\
%&=&
%\Phi \varepsilon_w(i \zeta_m)+(1-\Phi) \varepsilon_i(i \zeta_m),
%\end{eqnarray}
where $\Phi$ is the volume ratio of liquid water ($w$) and $1-\Phi$ is the volume ratio of solid ice ($i$). We can then use the parameterised dielectric function for imaginary frequencies for water and ice~\cite{LUENGOMARQUEZMacDowell2021}, and illustrate impacts due to the relative amounts of liquid water from $\Phi=$0.1 to 0.9 (in steps of 0.1).
% (check, 2023-02-02)

\par The interaction energy in a three-layer system is given by the well-known Casimir-Lifshitz free energy per unit area at temperature $T$, reading as~\citep{Dzya}
\begin{equation}
F(d) =\frac{k_{\rm B}T}{2 \pi}{\psum_{m=0}^\infty}  \int\limits_0^\infty \mathrm d k^\parallel \, k^\parallel \sum_{s=\rm{E,H}}  \ln\left(1-r^{s}_{21}r^{s}_{23}
 \mathrm e^{-2\kappa_2^\perp d}\right)\,, \label{100}
\end{equation}
where ${k}^\parallel$ is the modulus of wave vector ${\bf k}=(2\pi/\lambda)\hat{\bf k}$ parallel to the surface, and $\kappa_i^\perp({k}^\parallel,i \zeta_m)= \sqrt{{k^\parallel}^2+\varepsilon_i\zeta_m^2/c^2}$ with $i=1,2,3$ corresponding to the different materials in the structure and $\zeta_m=m\, 2\pi k_{\rm B}T/\hbar$ being the Matsubara frequencies, in which the reduced Planck constant $\hbar$, the Boltzmann constant $k_{\rm B}$, and the speed of light $c$, presents. The primed sum indicates the half-weighted $m=0$ term. The electromagnetic polarisations are transverse electric, $s=$E, and transverse magnetic, $s=$H. Reflections between the interfaces lead to
\begin{equation}
    r^{\rm E}_{ij}({k}^\parallel,i \zeta_m) = \frac{\kappa^\perp_i-\kappa^\perp_j}{\kappa^\perp_i+\kappa^\perp_j},\quad r^{\rm H}_{ij}({k}^\parallel,i \zeta_m) = \frac{\varepsilon_j\kappa^\perp_i-\varepsilon_i \kappa^\perp_j}{\varepsilon_j \kappa^\perp_i+\varepsilon_i \kappa^\perp_j} \,. \label{eq:rtTETM}
\end{equation}
This free energy $F(d)$ describes the interaction on the interface between the partially melted ice,  $\varepsilon_1(i \zeta_m)$, and the vapor,  $\varepsilon_3(i \zeta_m)$, across an ice (or water) film,  $\varepsilon_2(i \zeta_m)$, with thickness $d$.

\par It is useful to explore the case of thin ice (water) films in the non-retarded limit, where the frequency and spatial dependence separate.
In this case, the Casimir-Lifshitz free energy can be expressed in terms of the Hamaker constant~\cite{BERGSTROM1997125} ($A_{123}=-12 \pi d^2 F$),
\begin{equation}
A_{123}=\frac{-3 k_B T}{2} {\sum_{m=0}^\infty}^\prime \int_0^{\infty}{dx x}
 \ln\widehat{\Delta}_{123}(i \zeta_m;x)\,.
\label{eq:Hamaker}
\end{equation}
in which the function $\widehat{\Delta}_{123}(i \zeta_m;x)$ and the reduced reflection coefficient $\widehat{r}_{ij}$ are expressed as
\begin{eqnarray}
&&
\widehat{\Delta}_{123}(i \zeta_m;x)=1+\widehat{r}_{12}(i \zeta_m)\widehat{r}_{23}(i \zeta_m)e^{-x},\nonumber\\
&&
\widehat{r}_{ij}(i \zeta_m)= \frac{\varepsilon_j(i \zeta_m)-\varepsilon_i(i \zeta_m)}{\varepsilon_j(i \zeta_m)
+\varepsilon_i(i \zeta_m)}\,.
\end{eqnarray}

%\subsection*{Supercooled temperatures}

%At temperatures slightly below the triple point of water, the liquid fraction in the partially melted ice is in a supercooled state.  In the current work, we neglect this temperature dependence.

\subsection*{Results}
\par
\begin{figure}
  \centering
  \includegraphics[width=1.0\columnwidth]{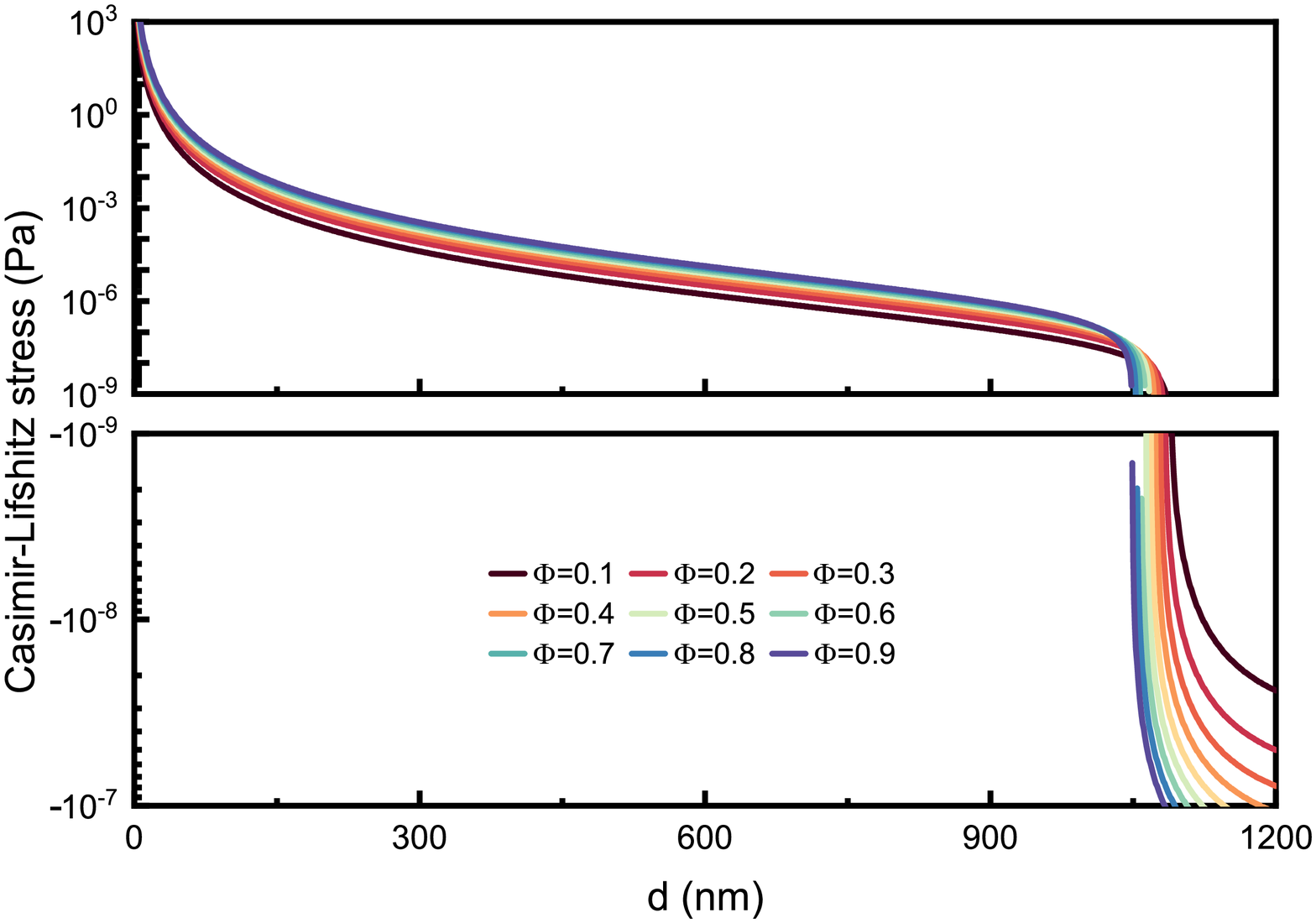}
  \caption{\label{fig:CasLifStress} Casimir-Lifshitz stress for a layered system with partially melted ice-ice-vapor, and  $\Phi=0.1,\ 0.2, \cdots, 0.9$. The zeros of stresses indicate the equilibria.
  }
\end{figure}

 As has been discussed~\cite{Ser2018}, the Hamaker constant reflects the short-range van der Waals interaction. At intermediate separations, the finite velocity of light reduces the short-ranged van der Waals interaction, as was first predicted by Casimir\,\cite{Casi}.
The $m=0$ Hamaker constant reflects the long-range  interaction~\cite{Dzya}.  Therefore, we can deduce from calculated Hamaker constants presented in Table\,\ref{tb1}, that the short-range repulsion and long-range attraction lead to finite growths of ice films on partially melted ice-water vapor interfaces. In contrast, for the case of a cold partially melted ice surface close to the melting temperature of ice, no water film growth is expected (except possible atomically thin liquid water layers originating from structural forces\,\cite{LuengoMarquez_IzquierdoRuiz_MacDowell2022}). We explore the detailed Casimir-Lifshitz stresses, derived from the corresponding free energy per unit area, for a few illustrative examples in Fig.\,\ref{fig:CasLifStress}.
The zero of Casimir-Lifshitz stress, corresponding to the free energy minimum here, is employed to identify the thickness of ice coating on the partially melted ice with a given liquid water volume fraction. Results in Table \,\ref{tb2} are obtained based on Fig.\,\ref{fig:CasLifStress}. The insensitivity of the equilibrium ice layer thickness to changes in $\Phi$, is due to the fact that $[\varepsilon_1(i \zeta_m)-\varepsilon_2(i \zeta_m)]=\Phi [\varepsilon_w(i \zeta_m)-\varepsilon_{i}(i \zeta_m)].$ Because of this scaling, the stability of the ice film, ultimately as the result of the energy minima, depends on the water content, but the equilibrium thickness is significantly influenced by retardation effects.
We emphasize\,\cite{Prachi2019role} that the quasi-planar theory is expected to give the same order of magnitude as the concentric sphere theory for ice growth in the size range we consider. The quasi-planar result will obviously approach closer to the exact value, as the initial core particle gets larger. We use this approximation in Table \,\ref{tb2} to estimate the relative volume growth for different mixed-phase ice/water nuclei of various sizes. We observe substantial volume growth ranging from $35\%$ and upwards.

\begin{table}
\centering
\begin{tabular}{c|c|c}
  \hline
  % after \\: \hline or \cline{col1-col2} \cline{col3-col4} ..
   $\Phi$ & $A_{1w3}$ ($A_{1w3;0}$) ($\rm meV$) & $A_{1i3}$ ($A_{1i3;0}$) ($\rm meV$)  \\
  \hline
   0.1 & $5.37\times10^1$ ($-2.85\times10^{-1}$) & $-4.90\times10^0$ ($ 3.12\times10^{-2}$) \\
 \hline
   0.2 & $4.76\times10^1$ ($-2.54\times10^{-1}$) & $-9.77\times10^0$ ($6.25\times10^{-2}$) \\
  \hline
   0.3 & $4.15\times10^1$ ($-2.23\times10^{-1}$) & $-1.46\times10^1$ ($9.40\times10^{-2}$) \\
    \hline
   0.4 & $3.54\times10^1$ ($-1.91\times10^{-1}$) & $-1.94\times10^1$ ($1.26\times10^{-1}$) \\
    \hline
   0.5 & $2.94\times10^1$ ($-1.60\times10^{-1}$) & $-2.42\times10^1$ ($1.57\times10^{-1}$)\\
    \hline
   0.6 & $2.35\times10^1$ ($-1.28\times10^{-1}$) & $-2.89\times10^1$ ($1.89\times10^{-1}$) \\
    \hline
   0.7 & $1.75\times10^1$ ($-9.63\times10^{-2}$) & $-3.36\times10^1$ ($2.21\times10^{-1}$) \\
    \hline
   0.8 & $1.17\times10^1$ ($-6.43\times10^{-2}$) & $-3.82\times10^1$ ($2.53\times10^{-1}$) \\
    \hline
   0.9 & $5.81\times10^0$ ($-3.22\times10^{-2}$) & $-4.28\times10^1$ ($2.86\times10^{-1}$) \\
   \hline
\end{tabular}
\caption{\label{tb1}The Hamaker constant $A_{123}$ and its contributions from the zeroth Matsubara term $A_{123;0}$ for various three-layer configurations with  partially melted ice (1) with different amount of water ($\Phi$) interacting with vapor (3). We consider two cases with the intermediate media (2) being either ice (i) or water (w).  }
\end{table}

 \begin{table}
\centering
\begin{tabular}{c|c|cc}
  \hline
  % after \\: \hline or \cline{col1-col2} \cline{col3-col4} ..
   a ($\,\mu\,m$) & $\Phi=0.1$\,\,($d_{2}\sim1.09\,\mu\,m$)& $\Phi=0.9$\,\,($d_{2}\sim1.05\,\mu\,m$) \\
    \hline
   1&  812\,\%&\,762\,\% \\
  \hline
   2 &  269\,\%&255\,\% \\
    \hline
   3 & 153 \,\%&146\,\% \\
  \hline
   4 & 106\,\%&101\,\% \\
     \hline
    5 & 81\,\%&77\,\% \\
    \hline
      6 & 65\,\%&62\,\% \\
       \hline
    7 & 54\,\%&52\,\% \\
    \hline
     8 & 47\,\%&45\,\% \\
    \hline
     9 & 41\,\%&39\,\% \\
    \hline
    10 & 36\,\%&35\,\% \\
    \hline
\end{tabular}
\caption{\label{tb2} The relative volume growth in percentage ($100\times[(a+d_{2})^3-a^3]/a^3 [\%]$) due to ice layer growth for different initial partially melted initial core with radius $a$.  We estimate\,\cite{Prachi2019role} the order of magnitude for the ice layer thicknesses using limiting results for large core radii. These estimated thicknesses are given by the zeros of the Casimir-Lifshitz stresses in Fig.\,\ref{fig:CasLifStress}. }
\end{table}

\subsection*{Discussions}
\par It is customary\,\cite{PruppacherKlett} to consider aerosol particles in the atmosphere with diameters  $r < 0.1\ \rm{\mu m}$, $0.1\ \rm{\mu m} \leq r \leq1.0\ \rm{\mu m}$ and $r > 1.0\ \rm{\mu m}$ as belonging to the nuclei mode, accumulation mode, and coarse mode, respectively. All these modes are exactly in the size ranges under study. Consider an illustrative example with a mixed phase droplet, i.e., a partially melted ice-water system, with initial core radius $a=1-10\ \rm{\mu m}$. For the case with $\Phi=0.1$, 90$\%$ of the volume is initially solid ice. Using the theory from the quasi-planar case, we predict that an ice layer about $1\ \rm{\mu m}$ thick forms, leading to the outer radii of $2-11\ \rm{\mu m}$ with $90\%$ ice in the inner core and $100\%$ ice in the shell. The estimated total volume increases in the now larger droplets are huge (about $800\%-30\%$), but it is important to recall that the estimates will be more reliable for the nucleus with a larger initial core radius. For a core radius of the order of $a$ = 1$\mu$m, we expect only an order of magnitude estimate with a reduction of the ice layer thickness from curvature effects. Thus, both the sizes of particles and the relative amount of frozen ice increase due to Casimir-Lifshitz interactions. This leads, in the larger perspective, to a secondary ice growth mechanism that is sufficiently important to have meteorological impacts. As demonstrated in the first half of the 20th century\,\cite{findeisen2015,StorelvmoTan2015} by Wegener, Bergeron, and Findeisen, mixtures of ice and water are thermodynamically unstable. Therefore, in subsequent steps, the inner, partially melted, ice core will eventually freeze, but end up with a substantially larger size, compared to the case when only the initial partially melted core had frozen.

\begin{acknowledgments}
The authors thank the "ENSEMBLE3 - Centre of Excellence for nanophotonics, advanced materials and novel crystal growth-based technologies" project (GA No. MAB/2020/14) carried out within the International Research Agendas programme of the Foundation for Polish Science co-financed by the European Union under the European Regional Development Fund, the European Union's Horizon 2020 research and innovation programme Teaming for Excellence (GA. No. 857543), and European Union's Horizon 2020 research and innovation programme (grant No. 869815) for support of this work.  We gratefully acknowledge Poland's high-performance computing infrastructure PLGrid (HPC Centers: ACK Cyfronet AGH) for providing computer facilities and support within computational grant no. PLG/2022/015458. We also thank the Terahertz Physics and Devices Group, Nanchang University for the strong computational facility support.
\end{acknowledgments}

\bibliography{PRL_drywet_bib}

\end{document}